\documentclass[twocolumn,showpacs,preprintnumbers,amsmath,amssymb,prb]{revtex4}
\usepackage{graphicx}
\usepackage{dcolumn}
\usepackage{color}
\usepackage{bm}
\begin{document}

\title{Structure, bonding and magnetism in cobalt clusters}
\author{S. Datta,$^1$ M. Kabir,$^{1,*}$ S. Ganguly,$^1$ B. Sanyal,$^2$ T. Saha-Dasgupta,$^1$ and A. Mookerjee$^{1,3}$}

\affiliation{$^1$Department of Material Sciences, S.N. Bose National Centre for Basic Sciences, JD Block, Sector-III, Salt Lake 
City, Kolkata 700 098, India \\ 
$^2$Theoretical Magnetism Group, Department of Physics, Uppsala University, Box 530, SE-75121  Uppsala, Sweden \\
$^3$Unit for Nanoscience and Technology, S.N. Bose National Centre for Basic Sciences, JD Block, Sector-III, Salt Lake 
City, Kolkata 700 098, India}

\date{\today}

\begin{abstract}
The structural, electronic and magnetic properties of Co$_n$ clusters ($n=$2$-$20) have been investigated 
using density functional theory within the pseudopotential plane wave method. An unusual hexagonal growth 
pattern has been observed in the intermediate
size range, $n=$15$-$20. The cobalt atoms are ferromagnetically ordered and the calculated magnetic moments 
are found to be 
higher than that of corresponding hcp bulk value, which are in good agreement with the recent 
Stern-Gerlach experiments. The 
average coordination number is found to dominate over the average bond length to determine the 
effective hybridization and consequently 
the cluster magnetic moment.
\end{abstract}
\pacs{75.75.+a, 36.40.Cg, 61.46.Bc, 73.22.-f}
\maketitle

\section{\label{sec:intro}Introduction}

 Study of finite size clusters is an important means of understanding how  
magnetic behavior evolves in reduced dimensionality. 
The 3$d$ transition metal (TM) elements are characterized by their unfilled $d$-shell, which gives
 rise to their magnetism and many other interesting physical and chemical properties.
Early transition metals are nonmagnetic in bulk solids, and only Fe, Co and Ni are 
known to be ferromagnetic among the 3$d$ metals. However, the small clusters of all the early 
transition metals are magnetic and those of late transition metals possess 
magnetic moments enhanced from their bulk values due to their spatial confinement. 
Local spin density based calculation\cite{11}  showed that the face centered cubic (fcc) phase is 
the lowest energy state for bulk Co in the paramagnetic phase, whereas the magnetic order stabilizes the 
hexagonal close packed (hcp) phase as the ground state.  This indicates a strong correlation between 
the stable structure and magnetism.  Although,   a metastable antiferromagnetic state exists for bulk Co,
 the ferromagnetic state is found to be the most stable for all crystal structures. This is unlike other 3$d$ transition metals Cr, Mn, and Fe which have a stable antiferromagnetic structure in their fcc phase. This 
means that not only the crystal structure, but also the electronic configuration controls magnetism. In the
 present communication, we  focus on Co$_n$ clusters to understand this interplay.

  The magnetic properties of bare Co$_n$ clusters were first investigated via Stern-Gerlach (SG)
 molecular beam deflection experiment by Bloomfield and co-workers for Co$_{20}$$-$Co$_{215}$ clusters\cite{1,doug} and by de Heer and co-workers for Co$_{30}$$-$Co$_{300}$ clusters.
 \cite{bill2,bill3,bill4} These studies showed that in the temperature range of 77$-$300 K, the Co$_n$ clusters display 
 high-field deflections, which  are characteristic of superparamagnetic behavior. 
 The superparamagnetic model for free clusters was recently revisited by Xu {\it et al.}\cite{co} and they proposed that adiabatic 
magnetization together with avoided Zeeman levels crossing in isolated clusters can lead to the same
high-field beam deflection behavior as observed in the  superparamagnetic spin relaxation. 
However, both the models 
predict the same high temperature limiting form for magnetization as given by the Curie law,\cite{co, markco}  $\langle M \rangle = \mu^2B/3kT$, where $\mu$ is the cluster magnetic moment, $B$ is the magnetic field and $T$ is the cluster temperature. 
 The intrinsic per-atom magnetic moment for small Co$_n$ clusters was found to be substantially
 larger than the bulk value\cite{1,doug,bill2, bill3, bill4, co, markco} and generally decreases with increasing cluster size,  eventually reaching the bulk value at $\sim$ 500 atoms.\cite{bill2} 
The enhancement in the magnetic moment in small clusters
 has been attributed to the lower coordination of the surface atoms resulting in a narrowing of the $d$-bands and hence greater spin polarization.

Information on the ground state geometry of the transition metal clusters is  usually obtained 
from the experiments involving chemical probe methods and photoelectron spectroscopy, though,
 such studies for the Co$_n$ clusters are very limited and not definitive. Reactions of Co$_n$ clusters 
with ammonia and water \cite{15} indicate icosahedral structures for the bare and ammoniated clusters in the 
size range $n=$ 50$-$120 and nonicosahedral packing for small (around 19 atoms) Co$_n$ clusters. Although the 
structures of ammoniated Fe$_n$, 
Co$_n$ and Ni$_n$ clusters in the size range of $n=$ 19$-$34 atoms have been found to be 
polyicosahedral,\cite{16} it has been mentioned that the bare clusters probably adopt
 a variety of structures. The photoionization experiment,\cite{17} indicated icosahedral atomic 
shell structures for large Ni$_n$ and Co$_n$ clusters of 50$-$800 atoms. However, structures were not
 well identified for small Co$_n$ clusters ($n\leqslant$50) because atomic sub-shell closings in different
 symmetry based clusters occur in close sequences.
These experimental results put together indicate that 
the icosahedral growth pattern for small sized Co$_n$ clusters is less evident. 

Theoretical works on cobalt clusters are limited and the available results are 
contradictory. Li and Gu\cite{8} performed first-principles calculation of small Co$_n$ clusters 
(4$\leqslant$$n$$\leqslant$19) using spin-polarized discrete variational method within local density functional
 theory (DFT). However, they had not optimized the structures and considered 
only some special structures with lattice parameters same as the bulk Co. Guevara {\it et al.}\cite{10}  used an unrestricted Hartree-Fock (HF) 
tight-binding formalism, starting from $spd$-bulk parameterization, but they only 
considered fixed body-centered cubic (bcc) and fcc geometries for a maximum of 177 atoms without 
structural relaxation. 
Andriotis and Menon\cite{9} have used a tight-binding  
molecular dynamics scheme to study cobalt clusters for some selected cluster sizes.  Castro {\it et al.}\cite{6}
 performed all-electron density functional calculations using both local density  
and generalized gradient approximations. However, the size  
of the clusters were  limited only up to 5 atoms. Recently, Lopez {\it et al.}\cite{7} studied Co$_n$ clusters
 (4$\leqslant$$n$$\leqslant$60), where minimization was done using an evolutive algorithm based on a many-body
 Gupta potential\cite{gp} and magnetic properties have been studied by a $spd$ tight-binding method.
 As compared to {\it ab-initio} methods, the parameterized tight-binding Hamiltonian
 reduces the computational cost drastically, but its main problem is the lack of transferability
 of its parameters. In particular, because of the lack of DFT like self-consistency the charge transfer effects are not 
properly accounted for and hence magnetic moment results are not fully reliable.

In this communication, we report the first-principles calculation of Co$_n$ clusters (2$\leqslant$$n$$\leqslant$20).
Without any symmetry constraints, we simultaneously relax the geometric and magnetic structure to find out the true ground state.
Our main interest is to study the evolution of structural, bonding and magnetic properties as a function of cluster size. We would also 
point out how the average bond length and coordination number determine the effective hybridization and hence the cluster 
magnetism.

\section{\label{sec:methodology} Computational Details}

The calculations are performed using density functional theory, within the pseudopotential
plane wave method.\cite{kresse2} We have used projector augmented wave (PAW) method
 \cite{blochl, kresse} and Perdew-Bruke-Ernzerhof (PBE)
exchange-correlation functional \cite{perdew} for spin-polarized
generalized gradient approximation (GGA).
The 3$d$ and 4$s$ electrons are treated as
valence electrons and the wave functions are expanded in the plane
wave basis set with the kinetic energy cut-off of 335 eV. Reciprocal
space integrations are carried out at the $\Gamma$ point. Symmetry
unrestricted geometry and spin optimizations are performed using
conjugate gradient and quasi-Newtonian methods until all the force
components are less than a threshold value of 0.005 eV/\r{A}. Simple
cubic super-cells are used with the periodic boundary conditions,
where two neighboring clusters are kept separated by at least 12 \r{A} 
vacuum space. This essentially makes the interaction between the cluster images 
negligible.
For each size, several initial geometrical structures have been
considered. To get the ground state magnetic moment we have explicitly
considered {\it all possible} spin multiplicities for each geometrical
structure. 
The binding energy per atom is calculated as,
\begin{equation}
E_b(\mbox{Co$_n$}) \ = \ \frac{1}{n}\left[ \rule{0mm}{4mm} \ n \ E(\mbox{Co}) \ - \ E(\mbox{Co$_n$}) \right],
\end{equation}
where $n$ is the size of the cluster. $E$(Co) and $E$(Co$_n$) are the total energies of isolated Co-atom and
$n$-atom Co$_n$ cluster, respectively. In such a definition, a positive sign in $E_b$ corresponds to binding.

\section{\label{sec:results}Results and discussions}

\subsection{\label{sec:Co2-10}Small Clusters: Co$_{2}$$-$Co$_{10}$}

Both experimental and theoretical predictions of the true ground state of the Co$_2$ dimer are 
controversial.
The first experimental estimation of Co$_2$ dimer bond length and binding energy has been made by mass
spectroscopy,\cite{19} which are 2.31 \r{A}  and 1.72 eV, respectively. 
However, more recent collision-induced dissociation (CID) experiment\cite{18} has estimated 
an upper bound of 1.32 eV to the dimer dissociation energy. 
The present calculation  gives dimer binding energy as 1.45 
eV/atom and a bond length of 1.96 \AA, which is 78\% of the bulk hcp Co. 
The Co atoms in dimer have bonding configuration closer to 3$d^8$4$s^1$ than that of the 
isolated Co atom, which is $3d^74s^2$ and 
in addition to the highly delocalized 4$s$ electrons, the more localized 3$d$ electrons 
also contribute strongly to the bonding,\cite{20} which consequently, produces a shorter bond length 
for the dimer.
Compared to the neutral Co$_2$ dimer, the experimentally\cite{18,cation1,cation2} predicted  
that the bonding in  Co$_2^+$ dimer cation is much strong, $\sim$ 2.73 $\pm$ 0.27 eV, which is formed by combining a neutral 
 Co ($3d^74s^2$) atom with a Co$^+$ ($3d^8$) cation in their respective ground state. Therefore, no promotional 
energy is required to form the cationic dimer. 
In fact, the bonding in Co$_2^+$ is relatively strong compared with other first-row 
transition metal dimer  cations. 
On the other hand, the formation of neutral Co$_2$ dimer requires 3$d^7$4$s^2$ $\rightarrow$ 3$d^8$4$s^1$ promotion \cite{note1}
 for both the
Co atoms, which is 0.42 eV.\cite{new}
We found that Co$_2$ dimer has a total magnetic moment of 4  $\mu_B$, which is also 
consistent with mass spectroscopic measurement\cite{19}.
Our estimates 
agree with the previous first-principles calculations.\cite{6,5,23,harris}

\begin{figure}[!t]	
\includegraphics[width=9cm,keepaspectratio]{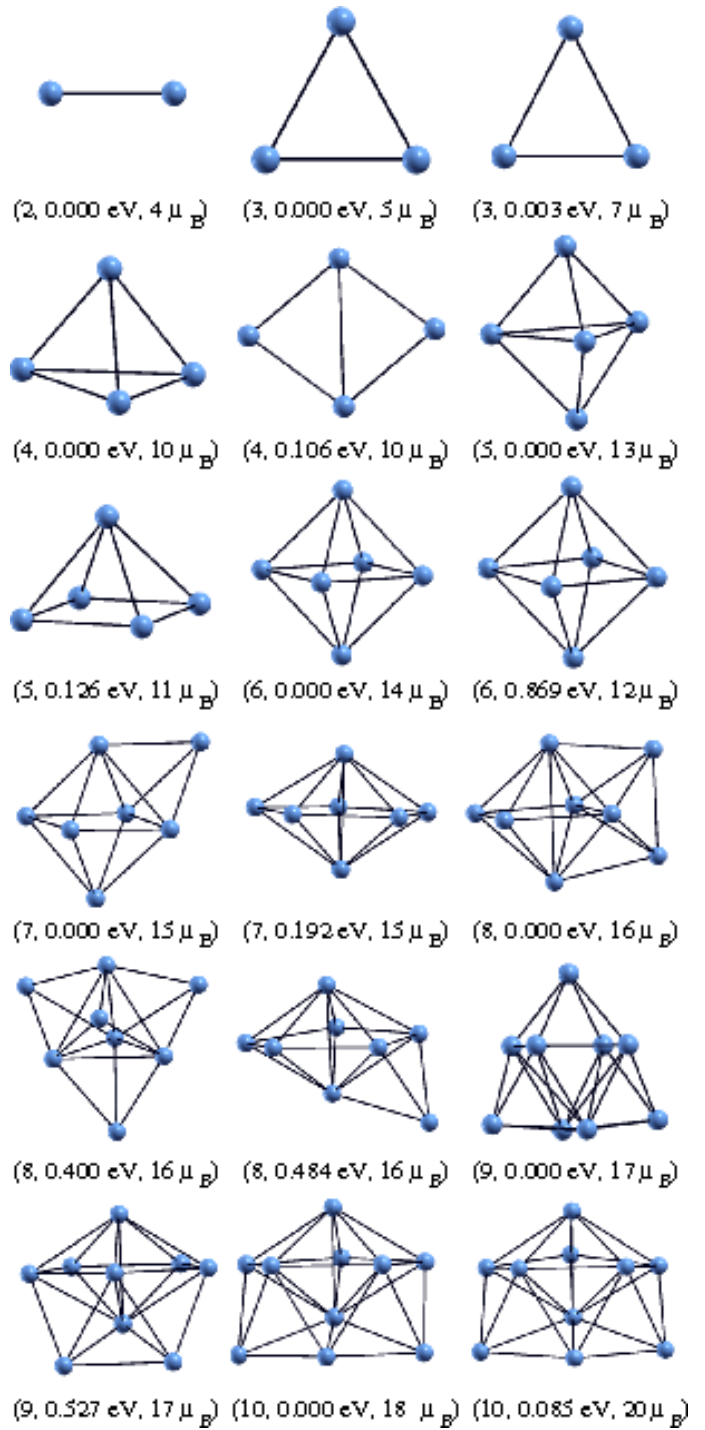}
\caption{\label{fig:1to10} (Color online) Equilibrium geometries of the two 
energetically lowest isomers of Co$_n$ clusters for $n =$ 2$-$10 (except $n=$ 8, for 
which three consecutive isomers have been shown). Numbers in 
the parenthesis represent number of atoms in the cluster, relative energy to 
the respective ground state and total magnetic moment, respectively.}
\end{figure}

In Fig.\ref{fig:1to10} we show the geometrical structures of Co$_n$ clusters for the ground state and the first
isomer for $n=$2--10 and the calculated binding energy, relative energy to the ground state and magnetic moment
are given in the Table \ref{tab:bemag} for the entire size range, $n=$2---20.
 
For Co$_3$ cluster, we have studied  both the linear and the triangular structures. 
An isosceles triangle with total magnetic moment 5 $\mu_B$ is 
found to be the ground state with binding energy 1.78 eV/atom. Each of the two equal sides has length 
of 2.19 \r{A} and other one has 2.10 \r{A} length. Another isosceles triangle with two long and one 
short bond lengths of 2.25 and 2.06 \r{A}, respectively, is found to be nearly degenerate with the
ground state structure (energy difference is only 3 meV). According to the present calculation, the linear
structure lies much higher in energy. The optimal linear structure has a total magnetic moment of 7 $\mu_B$ 
and lies 0.43 eV 
higher than the ground state.
Present result is consistent with the spin resonance spectra of Co$_3$ in Ar/Kr matrix, which 
indicated a triangular structure with a total moment of 5 or 7 $\mu_B$ to be the ground state.\cite{last}
The bonding in Co$_3^+$ cation is stronger than the neutral Co$_3$ as it is the case for dimer.\cite{18}
Yoshida {\it et al.}\cite{yoshida} reported that Co$_3^-$ has a linear structure with bond 
distance of 2.25$-$2.50 \r{A} based on their photoelectron spectroscopic study.
In agreement, previous all-electron (AE) density functional calculation\cite{6} predicted an isosceles triangle
(2.12, 2.12, 2.24 \AA) with a magnetic moment of 1.7 $\mu_B$/atom as the ground state for Co$_3$. This is also
consistent with the tight-binding study,\cite{9} but they predicted much higher bond lengths and magnetic moment.

%**************************************************************************************************************
Three different initial geometries have been considered for Co$_4$ cluster: tetrahedral, rectangular and linear.
A distorted tetrahedron with a total magnetic moment of 10 $\mu_B$  appears to be  the most stable structure. It 
has 2.27 eV/atom energy and has an average bond length of 2.34 \AA. 
Among the six sides of this tetrahedral ground state, two pairs have equal 
lengths of 2.14 \AA, while the third pair is much larger, 2.72 \AA. 
There is no experimental result available on the structure of neutral tetramer. However, 
Yoshida {\it et al.}\cite{yoshida} predicted a tetrahedral structure with a bond length of 2.25 $\pm$ 0.2 \AA \
as the ground state for Co$_4^-$ anion.
The initial rectangular structure 
becomes a rhombus after optimization, which also has 10 $\mu_B$ magnetic moment and lies 0.11 eV higher
in energy from the ground state. It is the next 
energetically favorable state, which has sides of length 2.14 \r{A} and two diagonals of 2.67 \r{A} 
and 3.35 \AA. Our results for Co$_4$ are consistent with previous calculations.\cite{6,7,8,9,10} 
Castro {\it et al.}\cite{6} predicted a strong Jahn-Teller distorted tetrahedral ground state with bond lengths almost equal 
to the present values.  
The distorted tetrahedral ground state structure is accomplished by a reduction of 
some inter-atomic distances (and the enlargement of other bonds) until some short equilibrium bond lengths
result for which there is a more effective participation of the (short-range) 3$d$-electrons 
in the bonds. In fact, in the distorted tetrahedron, there are some bonds (these are always on opposite TM$-$TM sides), 
which have lengths close to that of the dimer. These short bonds have high 3$d$ contributions and they are, therefore, 
the major source of increase of the bonding 
in the distorted structure. 
For both the ground state and the first isomer, the magnetic moment is found to be 2.5 $\mu_B$/atom.
The optimal linear structure is at a much higher energy than the ground state.
%***************************************************************************************************

We took trigonal bi-pyramid, square pyramid and two planar structures: two triangles connected through a 
vertex and a pentagon as the initial structures for Co$_5$ cluster.
The trigonal bi-pyramid with total magnetic moment 13 $\mu_B$ is found to be the most stable structure. 
This structure has   
2.55 eV/atom binding energy and 2.34 \r{A} average bond length.
In this ground state (Fig.\ref{fig:1to10}), there are two types of bond lengths: all the sides of the upper and lower triangular pyramids
are of same length and are smaller (2.18 \r{A}), while those of the interfacing planar triangle are much larger, 2.65 \r{A}. 
Another triangular bi-pyramid and a square pyramid with an equal magnetic moment of 11 $\mu_B$ are found to be the
degenerate first isomer. They lie 125 meV higher in energy. 
The optimal planar pentagon with 11 $\mu_B$ magnetic moment lies much, 1.04 eV, higher and the double triangle structure lies 
even higher in energy from 
the ground state. Present results are in agreement with the previous AE-GGA calculation,\cite{6} where they predicted the
same geometric structure with 2.28 \r{A} average bond length and 2 $\mu_B$/atom magnetic moment as the ground state. 
On the other hand,
the prediction\cite{7} of average bond length and magnetic moment using Gupta potential is much higher, though 
it predicted the same geometry.  
 
We have studied the capped trigonal bi-pyramid, octahedron and pentagonal pyramid to search the 
ground state for Co$_6$ cluster. From now on for the larger clusters, the planar structures have been discarded by intuition. 
An octahedral structure with 14 $\mu_B$ total magnetic moment is found to be the ground state.
An initial capped triangular bi-pyramidal structure relaxes to the octahedral ground state. 
Each side of this octahedral ground state is about 2.27 \r{A} and has a binding energy of 2.93 eV/atom.
Another slightly distorted octahedron with 12 $\mu_B$ moment appears as the first isomer. However, 
it is 0.87 eV higher compared to the ground state. 
In the present calculation, the optimal pentagonal pyramid lies much higher (1.7 eV) in energy compared 
to the ground state and also has 12 $\mu_B$ magnetic moment.
Present result is in agreement with previous theoretical studies,\cite{7,8,22,23} and  
the octahedral structure is generally accepted as the most stable structure for Co$_6$ cluster. 
However, photoelectron spectroscopic study\cite{yoshida} predicted a 
pentagonal pyramid with bond distances $\sim$ 2.75 $\pm$ 0.1 \r{A} to be the most probable 
structure for Co$_6^-$ anion cluster, i.e., the geometrical structure might strongly be correlated with the 
charged state of the cluster. 

For Co$_7$ cluster, we considered capped octahedron, pentagonal bi-pyramid and 
bi-capped triangular bi-pyramid. After simultaneous relaxation of both geometrical and magnetic structure,
the capped octahedron 
with 15 $\mu_B$ magnetic moment appears as the most stable structure. This structure has an
average bond length of 2.29 \r{A} and has 2.97 eV/atom binding energy. The experimentally measured 
magnetic moment, 2.36 $\pm$ 0.25 $\mu_B$/atom\cite{markco}, is little higher than our result. 
The optimal pentagonal 
bi-pyramid, which is a building block of icosahedral structure, has a total magnetic moment of 15 $\mu_B$, which lies 
0.19 eV higher in energy from the ground state.
This is the first isomer and it has an average bond length of 2.32 \AA. 
The optimal bi-capped triangular bi-pyramid has a total magnetic moment of 15  $\mu_B$ and  
lies 0.42 eV higher.  However, using Gupta potential, Lopez {\it et al.}\cite{7} predicted a pentagonal bi-pyramidal structure
as the ground state, 
and a capped octahedra as the first isomer. 

We have studied three different geometries for Co$_8$ cluster: bi-capped octahedron, capped pentagonal 
bi-pyramid and tri-capped triangular bi-pyramid. The bi-capped octahedron with 16  $\mu_B$ magnetic moment 
is found to be the most stable structure. This ground state has 3.07 eV/atom binding energy and 
an average bond length of 2.30 \AA. 
The  experimentally measured magnetic moment, 2.51 $\pm$ 0.15 $\mu_B$/atom,\cite{markco} is higher than the present value.
The optimal tri-capped triangular bi-pyramid and the optimal capped pentagonal bi-pyramid have an equal magnetic moment of 16 $\mu_B$
but lie 0.4 and 0.48 eV higher in energy respectively. They are the first and second isomers (see Fig.\ref{fig:1to10}).

%****************************************************************************************************************************

%================================== Figure 1 =============================================================
%=========================================================================================================

For the Co$_9$ cluster, we considered tri-capped octahedron and bi-capped pentagonal bi-pyramid as 
initial configurations.
A distorted tri-capped octahedron is found to be the most stable
structure with 3.14 eV/atom binding energy. This ground state structure has 17 $\mu_B$ total magnetic moment, which is smaller
than the experimental value of 2.38 $\pm$ 0.11 $\mu_B$/atom.\cite{markco} 
The optimal bi-capped pentagonal bi-pyramid has 17 $\mu_B$ magnetic moment and lies 0.53 eV higher in energy. This
structure is found to the first isomer.
The present results are not in good agreement with the semi-empirical predictions,\cite{7} where they predicted the bi-capped 
pentagonal bi-pyramid as the ground state with a relatively high magnetic moment.

%============================================ Table 2 ======================================================================== 
\begin{table*}[!t] 
\caption{\label{tab:bemag}Binding energy, relative energy to the theoretically computed minimum energy state 
($\triangle E=E-E_{\rm {min}}$) and magnetic moment for Co$_n$ ($n=$2$-$20) clusters.
Recent SG experimental results\cite{co,markco,markprivate,xuprivate} of the magnetic moment are shown for comparison.}
{\begin{tabular}{cccccccccccccccc} 
\hline
\hline
Cluster &  $E_b$       & $\triangle E$ & \multicolumn{2}{c} {Magnetic Moment} &  & & 
Cluster &  $E_b$       & $\triangle E$ & \multicolumn{2}{c} {Magnetic Moment} &  & \\
        & (eV/atom) &   (eV)        & \multicolumn{2}{c} {($\mu_B$/atom)}  &  & &
        & (eV/atom) &   (eV)        & \multicolumn{2}{c} {($\mu_B$/atom)}  &  &       \\
        &           &               & Theory      &  SG Exp.   &  & &
        &           &               & Theory      &  SG Exp.   &  &  \\
\hline
Co$_2$     &    1.452  &    0.000  &    2.00      &      $-$           &       &  &
Co$_{15}$  &    3.397  &    0.000  &    2.07      & 2.38$\pm$0.03\footnotemark[1] (2.09$\pm$0.04)\footnotemark[2] &      \\
Co$_3$     &    1.783  &    0.000  &    1.67      &      $-$           &       &  &
           &    3.393  &    0.046  &    1.93      &                    &      \\
           &    1.783  &    0.003  &    2.33      &                    &       &  &
           &    3.388  &    0.125  &    2.20      &                    &      \\
Co$_4$     &    2.274  &    0.000  &    2.50      &      $-$           &       &  &
           &    3.385  &    0.169  &    1.80      &                    &      \\
           &    2.248  &    0.106  &    2.50      &                    &       &  &
           &    3.385  &    0.171  &    1.93      &                    &      \\
Co$_5$     &    2.553  &    0.000  &    2.60      &      $-$           &       &  &
Co$_{16}$  &    3.458  &    0.000  &    2.13      & 2.53$\pm$0.04\footnotemark[1] (2.32$\pm$0.01)\footnotemark[2] &      \\
           &    2.528  &    0.125  &    2.20      &                    &       &  &
           &    3.458  &    0.005  &    2.00      &                    &      \\
Co$_6$     &    2.929  &    0.000  &    2.33      &       $-$          &       &  &
           &    3.445  &    0.208  &    1.88      &                    &      \\
           &    2.784  &    0.869  &    2.00      &                    &       &  &
           &    3.439  &    0.308  &    2.25      &                    &      \\
Co$_7$     &    2.971  &    0.000  &    2.14      & 2.36$\pm$0.25\footnotemark[1]  &       &  &
           &    3.438  &    0.319  &    1.88      &                    &      \\
           &    2.944  &    0.192  &    2.14      &                    &       &  &
Co$_{17}$  &    3.514  &    0.000  &    2.06      & 2.24$\pm$0.04\footnotemark[1] (2.19$\pm$0.02)\footnotemark[2] &      \\
Co$_{8}$   &    3.074  &    0.000  &    2.00      & 2.51$\pm$0.15\footnotemark[1]  &       &  &
           &    3.506  &    0.123  &    2.18      &                    &      \\
           &    3.024  &    0.400  &    2.00      &                    &       &  &
           &    3.504  &    0.167  &    1.94      &                    &       \\
           &    3.013  &    0.484  &    2.00      &                    &       &  &
           &    3.490  &    0.407  &    1.82      &                    &       \\
Co$_9$     &    3.143  &    0.000  &    1.89      & 2.38$\pm$0.11\footnotemark[1]  &       &  &
           &    3.466  &    0.812  &    2.06      &                    &       \\
           &    3.084  &    0.527  &    1.89      &                    &       &  &
Co$_{18}$  &    3.555  &    0.000  &    2.00      & 2.07$\pm$0.04\footnotemark[1] (2.37$\pm$0.07)\footnotemark[2] &       \\
Co$_{10}$  &    3.137  &    0.000  &    1.80      & 2.07$\pm$0.10\footnotemark[1]   &       &  &
           &    3.554  &    0.024  &    2.11      &                    &       \\
           &    3.128  &    0.085  &    2.00      &                    &        &  &
           &    3.544  &    0.194  &    1.89      &                    &        \\
Co$_{11}$  &    3.205  &    0.000  &    1.91      & 2.42$\pm$0.09\footnotemark[1]  &        &  &
           &    3.523  &    0.571  &    2.00      &                    &        \\
           &    3.203  &    0.016  &    1.91      &                    &        &  &
Co$_{19}$  &    3.607  &    0.000  &    2.05      & 2.21$\pm$0.03\footnotemark[1] (2.48$\pm$0.04)\footnotemark[2]  &        \\
Co$_{12}$  &    3.252  &    0.000  &    2.00      & 2.26$\pm$0.08\footnotemark[1] (2.21$\pm$0.01)\footnotemark[2] &         &  &
           &    3.597  &    0.174  &    1.95      &                    &        \\
           &    3.243  &    0.103  &    1.89      &                    &        &  &
           &    3.581  &    0.478  &    1.84      &                    &        \\
Co$_{13}$  &    3.279  &    0.000  &    1.92      & 2.30$\pm$0.07\footnotemark[1] (2.00$\pm$0.06)\footnotemark[2] &        &  &
           &    3.559  &    0.901  &    1.74      &                    &         \\
           &    3.268  &    0.140  &    2.08      &                    &         &  &
           &    3.546  &    1.158  &    2.16      &                    &         \\
           &    3.266  &    0.167  &    2.38      &                    &         &  &
           &    3.542  &    1.220  &    1.95      &                    &          \\
Co$_{14}$  &    3.323  &    0.000  &    2.00      & 2.29$\pm$0.06\footnotemark[1] (2.11$\pm$0.02)\footnotemark[2] &         &  &
Co$_{20}$  &    3.620  &    0.000  &    2.00      & 2.04$\pm$0.05\footnotemark[1] (2.36$\pm$0.02)\footnotemark[2] &          \\
           &    3.322  &    0.004  &    2.00      &                    &         &  &
           &    3.607  &    0.262  &    1.90      &                    &           \\
           &    3.322  &    0.005  &    1.71      &                    &           &  &
           &    3.588  &    0.634  &    1.80      &                    &           \\
           &    3.322  &    0.007  &    1.86      &                    &          &  &
           &    3.576  &    0.891  &    2.10      &                    &          \\
           &    3.320  &    0.008  &    2.14      &                    &         &  &
           &    3.565  &    1.103  &    1.90      &                    &          \\

\hline
\hline
\end{tabular} }
\footnotetext[1]{From Knickelbein (Ref.8 and Ref.35)}
\footnotetext[2]{From Xu {\it et al.} (Ref.7 and Ref.36)}
\end{table*}
%=====================================================================================================================================
 
Different tri-capped pentagonal bi-pyramid (TCPBP) structures along with different tetra-capped octahedral structures
were taken as initial structures for Co$_{10}$ cluster. 
A TCPBP structure with 18 $\mu_B$ total magnetic moment is found to be the ground state.
The calculated magnetic moment in the ground
state is  smaller as compared to the neighboring sizes, 
which is indeed the case in experiment (cf. Fig.\ref{fig:magmom} and will be discussed later). This is because of the fact that TCPBP is an icosahedral
fragment based on pentagonal bi-pyramid. This is different from the structural growth seen for Co$_6$$-$Co$_9$ clusters, where the ground state 
structures are all octahedral based.
For this TCPBP ground state, average coordination and average bond lengths are slightly higher and the competing
effect of these two makes the magnetic moment smaller than its neighboring clusters. 
Another TCPBP with total magnetic moment 20 $\mu_B$ lies
0.08 eV higher in energy compared to ground state is found to 
be the first isomer. 
The experimental magnetic moment (2.07 $\pm$ 0.10 $\mu_B$/atom\cite{markco}) is larger compared to the predicted
ground state but very close to the first isomer, which is energetically very close to the ground state.
Lopez {\it et al.}\cite{7} predicted the same TCPBP structure, but with much higher, 2.45 $\mu_B$/atom, magnetic moment as the ground state, whereas 
Guevara {\it et al.}\cite{10} predicted a fcc ground state with comparable, 2 $\mu_B$/atom, magnetic moment.

\subsection{\label{Co10-20}Intermediate size clusters: Co$_{11}$$-$Co$_{20}$}

%=================================== Figure 2 =================================================================================
\begin{figure*}[!t]
\includegraphics[width=14cm,keepaspectratio]{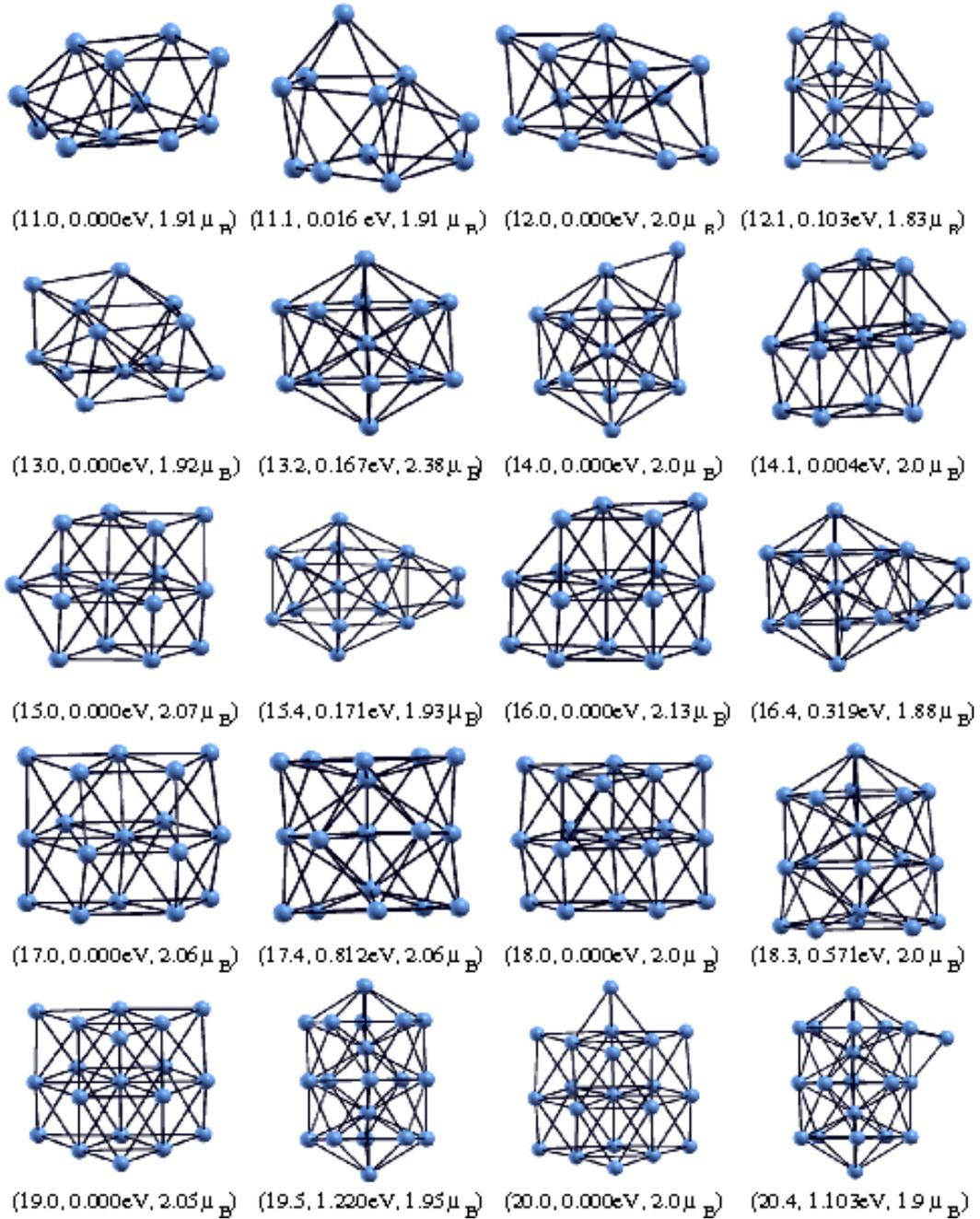}
\caption{(Color online) The ground state and a higher energy structure of Co$_n$
 clusters ($n=$11$-$20). For the size range $n=$13$-$20, we show the optimal icosahedra-derived structures for comparison. 
The first entry within the parenthesis, $n.k$, indicates that 
the structure corresponds to the $k$-th isomer of Co$_n$ cluster, second and third entries give the 
 relative energy to the ground state and magnetic moment per atom, respectively for the $k$-th isomer.
 $k=0$ corresponds to the ground sate.}
\label{fig:intermediate}
\end{figure*}
%==============================================================================================================================

With the increase in the number of atoms in the cluster, the determination of the ground state becomes 
a difficult task as the number of local minima in the potential energy surface increases very rapidly with the 
number of atoms in the cluster. In the search for the ground state structures, we have considered several possible
geometrical structures as initial guess and we relax all of them for {\it all possible} spin 
multiplicities for each $n$-atom cluster. The predicted minimum energy structure 
along with an higher energetic isomer for each cluster in the size range $n=$11$-$20 have been shown in Fig.\ref{fig:intermediate}.

The 13-atom hexagonal close packed structure consists of a hexagonal ring 
around a central atom and two triangular planes above and below it: 3,7,3 stacking, and 
the 13-atom close packed icosahedral structure has two pentagonal rings, two apex atoms
and a central atom: 1,5,1,5,1 stacking. 
The initial structures for both the Co$_{11}$ and Co$_{12}$ clusters,
have been derived from these 13-atom close packed structures by removing 1 or 2 atoms, respectively. 
As initial configurations for Co$_{11}$, we considered three hcp derived structures 
(4,7; 3,5,3 and 1,7,3 stacking) 
and two icosahedral derived structures (5,1,5 and 1,5,1,4 stacking). 
On relaxation, all the structures distorted heavily. The
distortion of the 4,7 hexagonal structure is such that one atom from the hexagonal
ring comes out of the plane (Fig.\ref{fig:intermediate}) and is the most stable configuration with binding 
energy 3.20 eV/atom. 
This structure has a total magnetic moment of 21 $\mu_B$, which is considerably smaller than that of the 
experimental value, 2.42 $\pm$ 0.09 $\mu_B$/atom.\cite{markco}  
The initial 1,5,1,4 icosahedral structure is found to be 
the next isomer after relaxation. This structure lies just 16 meV higher in energy from the ground state 
and also has 21 $\mu_B$ magnetic moment. 

For Co$_{12}$ cluster, we have considered two hcp structures: one with 3,6,3 stacking (without the central atom)
and the other with 2,7,3 stacking (with the central atom) and two icosahedral structures: a closed icosahedra
 without one apex atom and a closed icosahedra without the central atom. After relaxation the initial hcp 
structures undergo a considerable rearrangement, which consists of  a plane of 7 atoms coupled with another 
5-atoms plane (see Fig.\ref{fig:intermediate}). This structure having a total magnetic moment of 24 
$\mu_B$ is the lowest energy state, which has 3.25 eV/atom binding energy. A structural rearrangement has also been 
seen for 1,5,5,1 icosahedral structure, which looks like a hcp fragment of 3,5,4 stacking after relaxation. 
This structure has a total magnetic moment of 22 $\mu_B$ and has 3.24 eV/atom binding energy. The  
icosahedral structure with 1,5,1,5 stacking does not lead to such rearrangement after relaxation. However, 
the relaxed structure lies much higher in energy compared to the minimum energy state. The calculated  
magnetic moment in the lowest energy structure, agrees with the recent SG experiments\cite{co,markco} and the previous theoretical 
calculations.\cite{7,10}

We considered the icosahedral, hcp, cuboctahedral and fcc structures as initial structures for Co$_{13}$ 
cluster. A distorted hexagonal structure is found to be the minimum energy state 
and it looks like the most stable Co$_{11}$ structure along with  two additional capped atoms 
(see Fig.\ref{fig:intermediate}).
This structure has a total magnetic moment of 25 $\mu_B$ and has 3.28 eV/atom energy. 
Knickelbein found the experimental magnetic moment to be 2.30 $\pm$ 0.07 $\mu_B$/atom,\cite{markco} which is
slightly higher than the present value, 1.92 $\mu_B$/atom. However, this calculated value is in good
agreement with another recent SG experiment by Xu {\it et al.},\cite{co,xuprivate} which predicted 2.00$\pm$0.06 
$\mu_B$/atom moment.  
Another distorted hcp structure with total magnetic moment 27 $\mu_B$ is found to be the first isomer,
which lies 0.14 eV higher in energy from the ground state. The optimal icosahedral structure has a comparatively large magnetic moment, 31 $\mu_B$, and 
it lies 0.17 eV higher in energy being the second isomer. 
The optimal fcc and cuboctahedral structures are much higher in energy with respect to the minimum energy state. 
The present prediction of hcp structure as the minimum energy state is in agreement with the previous 
tight-binding prediction.\cite{9} However, there are some reports,\cite{7,22} which
favor the icosahedral structure as the ground state. 

For Co$_{14}$ cluster, the trial structures are complete icosahedra with a single atom capping and
a hexagonal structure with 3,7,4 stacking. The optimal icosahedral and the optimal hexagonal structures are 
found to be degenerate. The energy separation is only 4 meV. Both of these structures have equal (28 $\mu_B$) 
magnetic moment (see Fig.\ref{fig:intermediate}). We also found several isomers which lie very close
to these structures: An icosahedral structure (24 $\mu_B$), a hexagonal structure 
(26 $\mu_B$) and another icosahedra (30 $\mu_B$) lie only 5, 7 and 8 meV above the ground state, respectively 
(see Table \ref{tab:bemag}). The very recent SG experimental predictions range magnetic moment from 
2.11$\pm$ 0.02 $\mu_B$/atom\cite{co,xuprivate} to  2.29 $\pm$ 0.06 $\mu_B$/atom\cite{markco,markprivate}  for 
Co$_{14}$ cluster, which is in good agreement with the present result.

The Co$_{15}$ trial structure with lowest energy is the hexagonal structure with 4,7,4 atomic staking.
 This structure has 31 $\mu_B$ magnetic moment, which is 
in agreement with the SG experiment
of Xu {\it et al.}, which is 2.09$\pm$0.04 $\mu_B$/atom.\cite{co,xuprivate} However, 
Knickelbein predicted a larger value.\cite{markco,markprivate} The other hexagonal structures with 
total magnetic moments 29, 33 and 27  $\mu_B$ lie $\sim$ 0.05, 0.12 and 0.17 eV higher than the minimum energy state, 
respectively. The optimal icosahedral structure of 1,5,1,5,1,2 stacking (with 29 $\mu_B$ magnetic moment)
is the fourth isomer, which lies 0.17 eV above the lowest energy state.  

The same kind of structural growth is observed in the case of Co$_{16}$ cluster. The hexagonal structure 
with a total magnetic moment of 34 $\mu_B$ is found to be the lowest in energy. This structure has  4,7,5 
stacking and 3.46 eV/atom binding energy. This structure is nearly degenerate (5 meV lower) with 
another hexagonal structure, which has 32 $\mu_B$ magnetic moment. The next two isomers are also of 
same hexagonal motif, which have 30 and 36 $\mu_B$ magnetic moment and they lie 0.21 and 0.31 eV higher 
in energy with respect to the lowest energy state, respectively. The optimal icosahedral structures with 
1,5,1,5,1,3 stacking and with magnetic moments 30 and 32 $\mu_B$ lie 0.32 and 0.36 eV higher, 
respectively. The optimal hexagonal and icosahedral structures are shown in Fig.\ref{fig:intermediate}.
The other icosahedral structure with 5,1,5,4 stacking is found to be much higher in energy.  
 
The hexagonal structure with total magnetic moment 35 $\mu_B$ is the lowest energy state for Co$_{17}$ cluster. This 
structure has 5,7,5 stacking and 3.5 eV/atom binding energy. The calculated magnetic moment, 2.06 $\mu_B$/atom, 
is slightly smaller than that of predicted by both the recent experiments\cite{co,markco} (see Table 
\ref{tab:bemag}). The next three isomers also have hexagonal symmetry. They have 37, 33 and 31 $\mu_B$ 
magnetic moment and they lie 0.12, 0.17 and 0.41 eV higher than the lowest energy state, respectively. The optimal 
icosahedral structure (Fig.\ref{fig:intermediate}) has 5,1,5,1,5 stacking and lies much higher in energy. 

A 6,7,5-hcp trail structure (Fig.\ref{fig:intermediate}), which has a total magnetic moment 
of 36 $\mu_B$ is found to be the lowest energy state for Co$_{18}$ cluster. This structure has a binding 
energy of 3.61 eV/atom. The magnetic moment is in agreement with Knickelbein, 2.07$\pm$0.04 
$\mu_B$/atom.\cite{markco, markprivate} However, Xu {\it et al.} predicted a higher value, 
2.37$\pm$0.07 $\mu_B$/atom.\cite{co,xuprivate} Another two hcp structures with total magnetic 
moments 38 and 34 $\mu_B$ lie 0.02 and 0.19 eV higher in energy, respectively and are the first 
and second isomers. The optimal icosahedral structure has a total magnetic moment of 36 $\mu_B$, which 
lies much higher (0.57 eV) in energy and is the third isomer. 

For Co$_{19}$ cluster, we investigated a double icosahedral structure, a hcp structure with
6,7,6 stacking and a cuboctahedral structure. The hcp structure (Fig.\ref{fig:intermediate}) with  
39 $\mu_B$ magnetic moment appears as the most stable structure among all the trial structures.  The calculated 
magnetic moment, 2.05 $\mu_B$/atom, is closer to the value of Knickelbein, 2.21$\pm$0.03 
$\mu_B$/atom\cite{markco,markprivate} than that of the value predicted by Xu {\it et al.}, 2.48$\pm$0.04 
$\mu_B$/atom.\cite{co,xuprivate} The next four isomers are also found to be of same hcp 
packing. These isomers with total magnetic moments 37, 35, 33 and 41 $\mu_B$ lie 0.17, 0.48, 
0.90 and 1.16 eV higher than the minimum energy state, respectively. On the other hand, the optimal 
icosahedral structure (Fig.\ref{fig:intermediate}) has 37  $\mu_B$ magnetic moment and  lies 1.22 eV higher from 
the ground state. The fcc and hcp fragments have also been proposed as ground state structures in the previous theoretical 
calculations.\cite{8,10} Our results are in according with the predicted hcp structure. Also 
some calculations\cite{7,9} predicted icosahedral ground state for Co$_{19}$. 

Among all the considered structures the capped 19-atom hexagonal structure (Fig.\ref{fig:intermediate}) 
is found to be the lowest energy state 
for Co$_{20}$ cluster. The calculated magnetic moment is found
to be 2 $\mu_B$/atom for this structure, which is in agreement with the value predicted by Knickelbein
\cite{markco} (2.04$\pm$0.05 $\mu_B$/atom). However, the moment predicted by Xu {\it et al.} is much higher 
(2.36$\pm$0.02 $\mu_B$/atom\cite{co}). Similar to what we have seen for Co$_{15}$$-$Co$_{19}$ clusters, the
next few isomers are also of hexagonal motif. The hcp structures with total magnetic moments 38, 36, and 42
$\mu_B$, which are 0.26, 0.63 and 0.89 eV higher are found to be the first, second and third isomers, 
respectively. The optimal icosahedral structure has a total magnetic moment of 38 $\mu_B$ and appears as
the forth isomer (Fig.\ref{fig:intermediate}). However, this structure lies much higher in energy.

\subsection{\label{BE}Binding energy, stability and dissociation energy}

%======================================== Figure 3 =================================================================================
\begin{figure}[b]
\rotatebox{270}{\includegraphics[height=8.0cm,keepaspectratio]{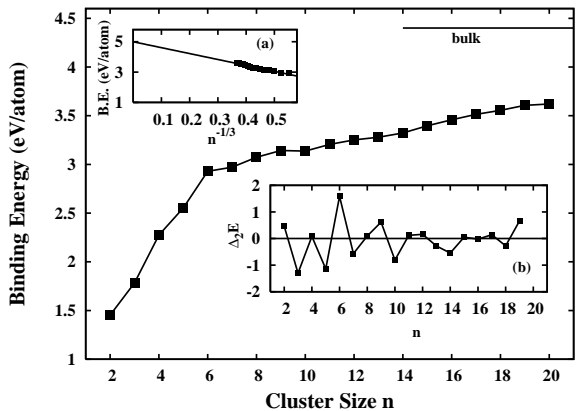}}
\caption{Plot of binding energy per atom as a function of cluster size $n$ for the ground state structures of the entire size 
range 2$\leqslant n \leqslant$20.  (a) Plot of binding energy per atom (B.E) as a function of $n^{-1/3}$ for the clusters Co$_n$, 
6$\leqslant n \leqslant$20 and a linear fit ($-$3.90 $n^{-1/3}$ + 5.00) to the data. (b) Plot of second difference in total 
energy ($\Delta_2 E$), which represents the relative stability.}
\label{fig:binding}
\end{figure}  
%===================================================================================================================================
%========================================= Figure 4 ================================================================================
\begin{figure}[t]
\rotatebox{270}{\includegraphics[height=8.5cm,keepaspectratio]{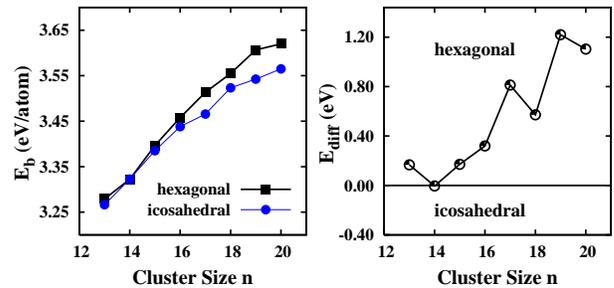}}
\caption{(color online) Plot of binding energy per atom ($E_b$) for optimal hexagonal and optimal icosahedral structures (left) 
and plot of total energy difference between these optimal structures, $E_{\rm {diff}}=-[E({\rm {hexagonal}}) - E(\rm {icosahedral})]$, 
(right) for the size range $n=$13$-$20. $E_{\rm {diff}}$ increases with cluster size.}
\label{fig:diff}
\end{figure}  
%===================================================================================================================================

%==============================Figure 5a ===========================================================================================
\begin{figure}[!b]
\rotatebox{270}{\includegraphics[height=8.5cm,keepaspectratio]{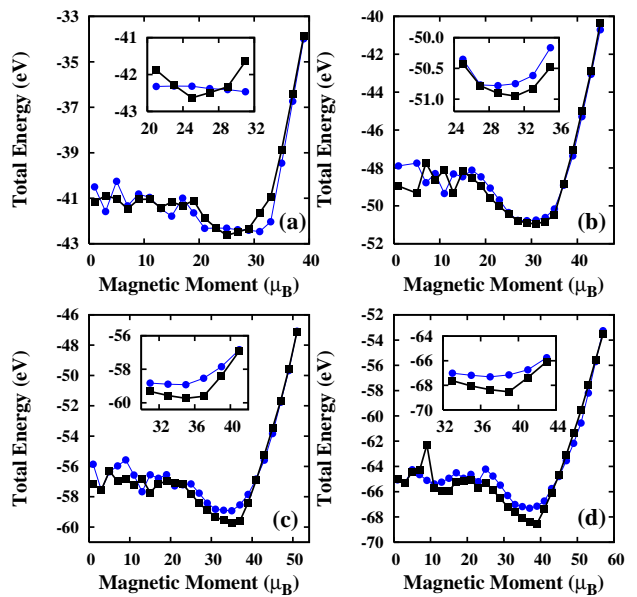}}
\caption{(color online) Plot of total energy as a function of magnetic moment for Co$_{13}$ (a), Co$_{15}$ (b),
Co$_{17}$ (c) and Co$_{19}$ (d) clusters.  The {\large{$\bullet$}} ({\tiny{$\blacksquare$}}) represents icosahedral (hexagonal) structure.
Insets represent magnification around the minima.}
\label{fig:energyvsmagneticmoment}
\end{figure}  
%===================================================================================================================================

Calculated binding energies are plotted in Fig.\ref{fig:binding} for the ground states of Co$_n$ 
clusters in the size range $n=$2$-$20.
Since the coordination number increases with the number of atoms in the cluster, the binding energy increases
 monotonically.
The binding energy of the largest cluster studied here (Co$_{20}$) is 3.62 eV/atom, which is about 82\% of the
 experimental bulk value, 4.4 eV/atom,\cite{kittelbook} for hcp Co. Upon extrapolation of the linear fit
 of the binding energy per atom data to $n^{-1/3}$ $\rightarrow$ 0 (Fig.\ref{fig:binding}(a)), we can estimate the 
binding energy of the infinitely large cluster. This is found to be 5.0 eV/atom, which is  larger than
 the experimental value for hcp bulk Co. However, within the same level of theory we found the hcp bulk cohesive
 energy to be 5.11 eV/atom, which is close to the extrapolated value but again larger than the experimental value.
 This overestimation is consistent with 
the DFT calculation.\cite{11} Calculated binding energies for the optimal hexagonal and 
optimal icosahedral structures (see Fig.\ref{fig:intermediate} for the optimal geometries) 
and the corresponding energy difference ($E_{\rm{diff}}$) between them are plotted in
Fig.\ref{fig:diff} for the size range $n=$13$-$20. The optimal hexagonal structures are always found to be the 
ground state for this size range except for Co$_{14}$, where the optimal hexagonal and icosahedral structures are
found to be degenerate. Moreover, in this size range, next few isomers are also of hexagonal motif and the optimal
icosahedral structures appear as higher energy (third, fourth or fifth) isomers for $n=$15$-$20 
(see Fig.\ref{fig:energyvsmagneticmoment}). The energy difference
between the hexagonal ground state and optimal icosahedral structures increases with increasing cluster size
making icosahedral structures more and more unfavorable. We plot the energy variation as a function
of cluster magnetic moment for icosahedral and hexagonal Co$_{13}$, Co$_{15}$, Co$_{17}$, and Co$_{19}$  
clusters in Fig.\ref{fig:energyvsmagneticmoment}. Both the structures show similar qualitative 
behavior for all the clusters and they have hexagonal minima around $\sim$ 2 $\mu_B$/atom moment.
   
We calculate the second difference in the total energy: 
\begin{equation}
\Delta_2 E(n) = E(n+1)+E(n-1)-2E(n), 
\end{equation}
where $E(n)$ represents the total energy of an
$n-$atom cluster. 
Calculated $\Delta_2E$ has been plotted in Fig.\ref{fig:binding}(b),
where we see the peaks at $n =$ 6 and 9, i.e., the clusters with 6 and 9 atoms are particularly 
more stable than their neighboring clusters. The stable structure for Co$_6$ is 
a octahedron and for Co$_9$, it is a distorted tri-capped octahedron. 
The CID experiment\cite{18} has also been indicated a maximum at $n$ = 6 in the measured dissociation energy, 
which 
indicates a higher stability of the hexamer. The extra stability of hexamer indicates that the octahedral structure 
can act as a building block for larger size clusters and, indeed, for Co$_{15}$$-$Co$_{20}$ clusters, we have
found a distinct hexagonal growth pattern and an octahedron is just a fragment of a hexagonal structure. 
The calculated stability (Fig.\ref{fig:binding} (b)) shows minima at $n$ = 3, 5, 7, 10 and 14, which 
are related to their weak bonding.

%========================================== Figure 5 ==========================================================================
\begin{figure}[!b]
\rotatebox{270}{\includegraphics[height=8.0cm,keepaspectratio]{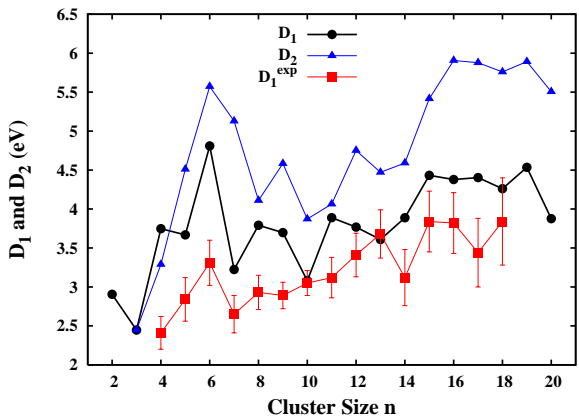}}
\caption{\label{fig:diss} (Color online) Plot of single channel, $D_1$ and dimer channel, $D_2$, dissociation energies as a 
function of cluster size $n$ for the GS configurations. We compare our calculated single channel dissociation with the CID 
experimental result.\cite{18}}
\end{figure}
%==============================================================================================================================

This can be further demonstrated by studying the dissociation energies as an $n$-atom cluster fragments into 
$m$ and ($n-m$)-atom clusters. The $m$-channel dissociation energy can be calculated as, 
\begin{equation}
D_m(n) = E(m) + E(n-m) - E(n),
\end{equation}
where $E(n)$, $E(m)$ and $E(n-m)$ are the total energies of $n$, $m$ and $(n-m)$ atom clusters, respectively. 
We have plotted the calculated single channel ($D_1$) and dimer channel ($D_2$) dissociation energies in 
Fig.\ref{fig:diss} and $D_1$ is compared with the CID experiment by Hales {\it et al.}.\cite{18}
However, they have estimated this dissociation
energy through an indirect method: actually, they measured the single channel dissociation energy of Co$_n^+$ cation
cluster and derived the same for the neutral one by using   
the ionization
energies (IE) of the neutral clusters measured by Yang and Knickelbein,\cite{yang1} and 
Parks {\it et al.},\cite{parksco} i.e., 
\begin{equation}
D_1^{\rm{exp}} = D_1({\rm{Co}}_n^+)+IE({\rm{Co}}_n)-IE({\rm{Co}}_{n-1}). 
\end{equation}
The calculated single channel dissociation energy, $D_1$, shows a high peak at $n=$6 and dips at $n=$ 5, 7 and 10, which
are consistent with our stability analysis. However, we do not find any dip in the calculated dissociation energy at
$n=$14, as has been seen in the CID experiment. Generally, the single channel dissociation energy is the most
favorable except for $n=$4, where the dimer dissociation (Co$_4$ $\rightarrow$ Co$_2$ + Co$_2$) is more favorable
than the single channel (Co$_4$ $\rightarrow$ Co$_3$ + Co) dissociation. Table II shows the theoretically computed
single channel bond dissociation energy compared to the experimentally measured values \cite{18} for the entire range of clusters
having sizes $2$ to $20$.
 
\begin{table}[!t]
\caption{\label{tab:bde}Theoretically calculated single channel bond dissociation energies(BDE) compared with experimentally measured values \cite{18} for Co$_n$ ($n=$2$-$20). Experimental uncertainties are within parentheses. }
{\begin{tabular}{ccccccccc}
\hline
\hline
Co$_n$& & \multicolumn{2}{c} {BDE(eV)}&  & Co$_n$ & & \multicolumn{2}{c} {BDE(eV)}    \\
      & & Theory      &  CID Exp.   &  &        & & Theory      &  CID Exp.  \\
\hline
2     & & 2.90     & $\le$1.32   &  &  12    & &  3.77       & 3.41(0.28)  \\
3     & & 2.45     & $\ge$1.45   &  &  13    & &  3.61       & 3.68(0.31)  \\
4     & & 3.75     & 2.41(0.21)  &  &  14    & &  3.89       & 3.12(0.36)  \\
5     & & 3.67     & 2.84(0.28)  &  &  15    & &  4.43       & 3.84(0.39)  \\
6     & & 4.81     & 3.31(0.29)  &  &  16    & &  4.38       & 3.82(0.39)  \\
7     & & 3.22     & 2.65(0.24)  &  &  17    & &  4.40       & 3.44(0.44)  \\
8     & & 3.79     & 2.93(0.22)  &  &  18    & &  4.26       & 3.84(0.56)  \\
9     & & 3.70     & 2.89(0.17)  &  &  19    & &  4.54       &              \\
10    & & 3.08     & 3.05(0.16)  &  &  20    & &  3.88       &               \\
11    & & 3.89     & 3.12(0.26)  &  &        & &             &             \\
                                                                                         
\hline
\hline
\end{tabular} }
\end{table}

%========================================== Figure 6 ==========================================================================
\begin{figure}[t]
\rotatebox{270}{\includegraphics[width=8.0cm,keepaspectratio]{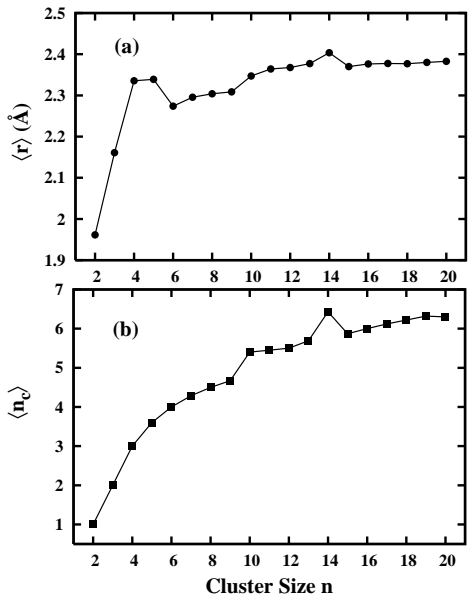}}
\caption{ Plot of (a) average bond length $\langle r \rangle$ and (b) average coordination $\langle n_c \rangle$, as function 
of cluster size $n$ for the ground state geometries. Dots and squares represent the data points, while solid line through them 
is guide to eye.}
\label{fig:blength}
\end{figure}
%==============================================================================================================================

To understand the optimized structures, we calculated the average bond lengths and average coordination number for the ground state geometries and 
plotted them in  Fig.\ref{fig:blength}(a) and Fig.\ref{fig:blength}(b), respectively, as a function of cluster size. These two quantities are closely related to the structure of the cluster. We define the average bond length as 
$\langle r \rangle$  = $\frac{1}{n_b}\sum_{i>j}r_{ij} $, where $r_{ij}$
is the bond distance between the $j$-th and $i$-th atoms, and $n_b$ is the number of such 
bonds. Here we consider that two atoms are bonded if their inter atomic distance
is within 2.91 \r{A}, which is around the average of the first (2.51 \r{A}) and  second (3.54 \r{A}) 
nearest-neighbor distances in bulk Co. 
The average coordination number in a cluster is defined as $\langle n_c \rangle$ = $\frac{1}{n}\sum_{k} n_k$ 
where $n_k$ is the number of neighbors within the chosen cut-off of the $k$-th
atom in the cluster of $n$ atoms. 
The convergence of the average bond length to the bulk value (2.51 \r{A}) is much faster than the 
convergence of average coordination, which is far below the bulk value (12 for hcp Co). 
Dips at $n =$ 6 and  9 in Fig. \ref{fig:blength}(a) indicate that the atoms in these clusters are closely spaced and strongly bonded compared to the neighbors,  and therefore are more stable than the neighboring structures. 
While the peaks at $n=$ 5, 10 and 14 in Fig.\ref{fig:blength}(a) and at $n=$10 and 14 in Fig.\ref{fig:blength}(b) indicate
that atoms in these clusters are far apart and slightly more coordinated then their neighbors,\cite{co14}
which results in a weak bonding in these clusters compared to their neighbors.

\subsection{\label{sec:mag}Magnetic moment}

\begin{figure}[t]
\rotatebox{270}{\includegraphics[height=8.0cm,keepaspectratio]{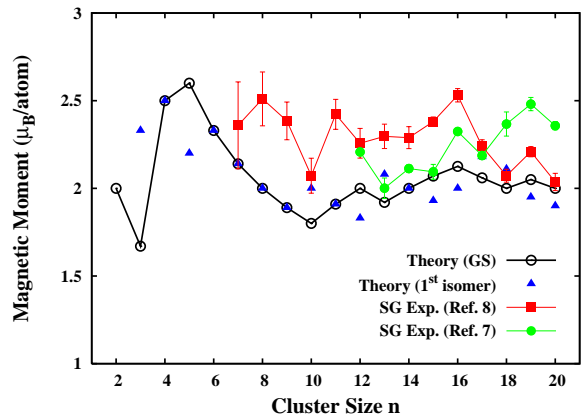}}
\caption{\label{fig:magmom} (Color online) Size dependent variation of magnetic moment of the corresponding ground states. 
Calculated magnetic moments of the first isomers have also been shown. Calculated magnetic moments are compared with 
experimental results (Ref.8 and Ref.7).}
\end{figure}

The calculated magnetic moments are plotted in Fig.\ref{fig:magmom} as a function of cluster size ($n$).
The Co$-$Co interaction is always ferromagnetic for the entire size range studied, as it is for hcp
bulk Co. However, the magnetic moment (2$-$2.5 $\mu_B$/atom) is larger than the hcp bulk value,
1.72 $\mu_B$/atom.\cite{kittelbook} This enhancement in moment for a few atom cluster can readily be 
understood from the more localized $d$-electrons resulting from the decrease in effective hybridization. 
The calculated magnetic moments are in fair agreement with the very recent SG experiments by 
Xu {\it et al.}\cite{co} and Knickelbein.\cite{markco} Fig.\ref{fig:magmom} shows a qualitative agreement 
between the calculated and the experimental values,\cite{co,markco,markprivate,xuprivate} though the calculated 
moments are always underestimated systematically. However, calculated moments are close to that of predicted by 
Xu {\it et al.}\cite{co,xuprivate} for the size range $n=$13$-$17 and in the size range $n=$18$-$20 they are 
close to the values predicted by Knickelbein.\cite{markco,markprivate} The underestimation of calculated moment 
may be due to the fact that we did not include spin-orbit interaction in the present calculation. Moreover,
one should remember that the magnetic moments in a magnetic deflection measurement are always derived assuming 
a model, which may influence the outcome. For example, Knickelbein\cite{markco} used either superparamagnetic 
or locked moment model, whether Xu {\it et al.}\cite{co} assumed an adiabatic magnetization model to derive the 
moments experimentally for cobalt clusters. It is important to note that both experiments show same size evolution
in general but there are some systematic differences. However, this is not due to the adoption of different 
models to calculate the magnetic moment as both the models resemble with the same Curie law for magnetization,\cite{co,
markco} but may be due to differing isomer distribution in the SG beam. 

The magnetic moment is strongly correlated with the effective hybridization, which is closely related to the 
average bond length  $\langle r \rangle$ and the average coordination number $\langle n_c \rangle$. As  $\langle n_c \rangle$
 decreases the magnetic moment should increase through the decrease in effective hybridization. On the other hand, the 
dependency of magnetic moment on $\langle r \rangle$ is directly proportional: a decrease in $\langle r \rangle$ 
results in decrease in magnetic moment through the enhancement in 
effective bonding. Fig.\ref{fig:blength}(a) and Fig.\ref{fig:blength}(b) show that as we go from $n=$ 4 to 
$n=$ 10, both $\langle r \rangle$ and $\langle n_c \rangle$ increase, whereas 
Fig.\ref{fig:magmom} shows that the magnetic moment per atom decreases. Therefore, between these 
two competing contributions ($\langle r \rangle$ and $\langle n_c \rangle$) to the magnetic
moment, the average coordination number dominates over the average bond length in the size range $n=$4$-$10.

%======================================== Figure 7 =================================================================================
%=================================================================================================================================== 

\begin{figure}[t]
\rotatebox{270}{\includegraphics[height=8.0cm,keepaspectratio]{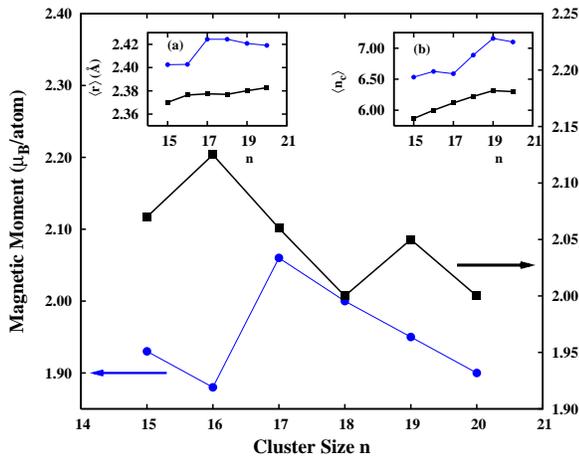}}
\caption{\label{fig:comp} (Color online) Comparison of magnetic moment between optimized hcp and optimized icosahedral structures for 
Co$_n$ in the size range $n=$15$-$20. The filled squares and filled circles correspond to results for hcp and icosahedral structures, 
respectively. The insets show the corresponding comparisons for (a) average bond length and (b) average coordination number.}
\end{figure}

In the intermediate size range, $n=$ 11$-$20, the variation of $\langle r \rangle$ (Fig.\ref{fig:blength}a) and $\langle n_c \rangle$ 
(Fig.\ref{fig:blength}b) is much slower with $n$, and therefore, the magnetic moment per atom does not vary rapidly. It is around 
2 $\mu_B$/atom for all the clusters in this size range. So, in this size range, it is hard to predict the dominant parameter for 
magnetism. To illustrate the effect of $\langle r \rangle$ and $\langle n_c \rangle$ on the magnetism in the intermediate size range 
we compare these two quantities for the optimal hcp and icosahedral structures (see insets (a) and (b) of Fig.\ref{fig:comp}). It is 
seen that for a hcp structure, both the $\langle r \rangle$ and $\langle n_c \rangle$ are smaller than those of corresponding 
icosahedral structure for a particular $n$-atom cluster. In addition the magnetic moments of optimal hcp clusters are always larger 
than or equal to that of the corresponding optimal icosahedral clusters (see Fig.\ref{fig:comp}), which again demonstrates that in 
this intermediate size range also the coordination dominates over the average bond length in deciding magnetism.  

%======================================== Figure 8 ==================================================================================
%====================================================================================================================================

\section{\label{summary}Summary and Conclusions}

We have systematically studied the structure, bonding and magnetism in Co$_n$ clusters in the size range
$n=$2$-$20. In the intermediate size range, the clusters adopt hcp structural packing, which is different from the
trend observed for the other 3$d$ transition metal clusters.\cite{thesis,MyMn}
In the size range $n=$15$-$20, the energy difference between the hexagonal minimum energy states
and optimal icosahedral structures increases with cluster size making the icosahedral structures more and 
more unfavorable with increasing size.
 The calculated magnetic
moments are in good agreement with both the recent SG experiments. It is found that the effect of 
average coordination number always dominates over the average
bond length to determine the effective hybridization and therefore, the magnetic moment of the clusters.

\acknowledgments
We thank Mark B. Knickelbein and X. Xu for providing their experimental magnetic moments.  
This work was done under the  Indian Department of Science and Technology (DST) Grant No. SR/S2/CMP-25/2003 and partly under 
the Asian-Swedish research links program.  S. D. is thankful to CSIR for financial support. T.S.D. thanks DST for Swarnajayanti fellowship.


\begin{thebibliography}{99}
\bibitem{note1}The 3$d^7$4$s^2$ $\rightarrow$ 3$d^8$4$s^1$ transition energy for a neutral cobalt atom is found to be
0.12 eV within our scheme of calculation.
\bibitem{co14} For $n =$ 14, two degenerate ground states have been observed: one with icosahedral symmetry and
other has hexagonal symmetry. The icosahedral structure is just slightly lower in energy compared to hexagonal
structure and for Fig.\ref{fig:blength}, we have taken this icosahedron as the ground state, for which both  $\langle r \rangle$ and  $\langle n_c \rangle$ are higher than the neighbouring hcp ground states.
\end{thebibliography}
\end{document}